\documentclass[doublecol]{epl2}
\title{Biconical critical dynamics}
\shorttitle{Biconical critical dynamics} %Insert here a short version of the title if it exceeds 70 characters

\author{R. Folk\inst{1} \and Yu. Holovatch\inst{1,2} \and G. Moser\inst{3}}
\shortauthor{R. Folk \etal}

\institute{
  \inst{1} Institute for Theoretical Physics, Johannes Kepler
University Linz, Altenbergerstrasse 69, A-4040, Linz, Austria\\
  \inst{2} Institute for Condensed Matter Physics, National
Academy of Sciences of Ukraine, Svientsitskii Str. 1, UA--79011
Lviv, Ukraine\\
 \inst{3} Department for Material Research and Physics, Paris Lodron University
Salzburg, Hellbrunnerstrasse 34, A-5020 Salzburg, Austria
}

\abstract{A complete two loop renormalization group calculation of
the multicritical dynamics at a tetracritical or bicritical point in
three-dimensional
anisotropic antiferromagnets in an external magnetic field is
performed. Although strong scaling for the two order parameters
(OPs) perpendicular  and parallel to the field is restored as found
earlier, in the experimentally accessible region the effective
dynamical exponents for the relaxation of the OPs remain different
since their equal asymptotic values are not reached.}

\begin{document}

\maketitle

Systems with more than one order parameter (OP) exhibit a rich
variety of phases separated by transition lines which might meet in
multicritical points. The interaction might favor simultaneously
ordering of two OPs. Such a doubled ordered phase is known as
supersolid phase \cite{liufisher72} and  is under investigation
since its possible observance in $^4$He \cite{chang04}. 
The  two OPs might describe the order in physically different phases: E.g.
from the normal fluid to the superfluid or to the normal solid phase. If 
both orders appears one might have transitions from the superfluid to the supersolid 
phase and from the normal solid phase to the supersolid phase. Another example would be
a system with transitions to a  superconducting and a magnetically 
ordered phase and a phase where both orderings appear. In purely magnetic systems 
the phases are characterized by different orderings in spin space.
There is a correspondence between the quantum liquid system and magnetic
systems where the supersolid phase corresponds to the biconical
phase \cite{matsuda70}. The existence of a biconical phase leads to
the occurrence of a tetracritical point where four second order
phase transition lines meet and which belongs to a new universality
class \cite{knf76}.

In the case of the three-component ($n=3$) three-dimensional ($d=3$)
anisotropic antiferromagnets in an external magnetic field in $z$
direction the disordered (paramagnetic) phase is separated from the
ordered phases by two second order phase transition lines: (i) one
to the spin flop phase (ordering in the spin space perpendicular 
to the external magnetic field) and (ii) one to the antiferromagnetic phase
(ordering parallel to the external field).
The point where these two lines meet is a multicritical point which
turned out to be either tetracritical or bicritical depending on
whether  the ordered phases are separated by an intermediate
biconical phase or not. The static phase transitions on each of the phase
transition lines belong for (i) to an XY-model with $n=2$ and for
(ii) to an Ising model with $n=1$ \cite{knf76}. Whether a bicritical or tetracritical
point is realized depends on the specific fourth order couplings \cite{knf76,partI}. 
Several antiferromagnets have been suggested for observing the theoretically 
proposed phase diagrams, a review on the experimental situation can be found 
in \cite{shapira}-\cite{aharony06}.

Concerning the
dynamical universality classes they might be different for the
systems mentioned above depending on the possible reversible and non reversible 
terms in the equations of motion and the conservation properties.
In the magnetic system considered here 
the transition (i) belongs to the
class described by model F  and (ii) belongs to the model C class
(for the definitions of the models see \cite{review}). At the
multicritical point the critical behavior is described by a new
universality class both in statics and dynamics characterized by the
biconical fixed point \cite{partI}. The advantageous feature of
these systems is that all the different OPs characterizing the
ordered phase are physically accessible. This is most important for
the dynamical behavior since the only other example belonging to
model F is the superfluid transition in $^4$He where the OP is  not
directly measurable. Here the OPs are the components of the
staggered magnetization. Their correlations (static and dynamical)
are experimentally accessible by neutron scattering. Realistic
models might be more complicated (see e.g. \cite{seabra10}) but the
behavior near the multicritical point is well described by the
renormalization group (RG) theory.

The dynamical model we analyze goes beyond the pure relaxational
dynamics \cite{partII} and has been considered by means of the field
theoretical RG approach in \cite{dohmjanssen77,dohmKFA,dohmmulti83}
replacing earlier mode coupling theories \cite{huber7476}. It was
argued that due to nonanalytic terms in $\epsilon=4-d$ a dynamical
fixed point (FP) in two loop order (which was calculated only
partly) qualitatively different from the one loop FP is found. In one
loop order the relaxation times of the components of the staggered
magnetization parallel and perpendicular to the external magnetic
field  scale  differently whereas in two loop order they would scale
similarly if the new FP would be stable. In addition it turned out
that the FP value of the timescale ratio of the two OPs cannot be
found by $\epsilon$ expansion and might be very small at $d=3$,
namely of ${\cal O}(10^{-86})$. A basic assumption of the above
analysis  was that within statics the Heisenberg FP is stable.
However it turned out in two loop statics using resummation
techniques that in $d=3$  the Heisenberg FP interchanges its
stability with the biconical FP \cite{partI}.  Here we  calculate
the complete functions in two loop order which allows us to consider
the non-asymptotic behavior near the multicritical point.

The non-conserved OP in an isotropic antiferromagnet is given by the
three-component vector $\vec{\phi}_0$ of the staggered
magnetization, which is the difference of two sublattice
magnetizations. In an external magnetic field applied to the
anisotropic antiferromagnet the OP splits into two OPs,
$\vec{\phi}_{\perp 0}=\left(\begin{array}{cc} \phi_0^x, & \phi_0^y
\end{array}\right)$ perpendicular to the field, and $\phi_{\|
0}=\phi_0^z$ parallel to the external field. In addition to the two
OPs the $z$-component of the magnetization, which is the sum of the
two sublattice magnetizations, has to be considered as conserved
secondary density $m_0$. The static critical behavior of the system
is described by the functional
\begin{eqnarray}\label{1}
{\cal H}\!=\!\int\!
d^dx\Bigg\{\frac{1}{2}\mathring{r}_\perp\vec{\phi}_{\perp 0}
\cdot\vec{\phi}_{\perp
0}+\frac{1}{2}\sum_{i=1}^{d}\nabla_i\vec{\phi}_{\perp 0}\cdot
\nabla_i\vec{\phi}_{\perp 0}
\nonumber \\
+\frac{1}{2}\mathring{r}_\|{\phi}_{\| 0} {\phi}_{\|
0}+\frac{1}{2}\sum_{i=1}^{d}\nabla_i\phi_{\| 0} \nabla_i\phi_{\| 0}
+\frac{\mathring{u}_\perp}{4!}\Big(\vec{\phi}_{\perp
0}\cdot\vec{\phi}_{\perp 0}\Big)^2 \nonumber \\
+\frac{\mathring{u}_\|}{4!}\Big(\phi_{\| 0}\phi_{\| 0}\Big)^2
+\frac{2\mathring{u}_\times}{4!}\Big(\vec{\phi}_{\perp
0}\cdot\vec{\phi}_{\perp 0}\Big) \Big(\phi_{\|
0}\phi_{\| 0}\Big) \Bigg\}  \\
+\frac{1}{2}m_0^2+\frac{1}{2}\mathring{\gamma_\perp}m_0\vec{\phi}_{\perp
0}\cdot\vec{\phi}_{\perp 0}
+\frac{1}{2}\mathring{\gamma_\|}m_0\phi_{\| 0}\phi_{\| 0}
-\mathring{h}m_0\Bigg\}  \ , \nonumber
\end{eqnarray}
with familiar notations for bare couplings
$\{\mathring{u},\mathring{\gamma}\}$, masses $\{\mathring{r}\}$ and
field $\mathring{h}$ \cite{partI,partII}. One may switch from the description of
real OP components
$\vec{\phi}_{\perp 0}$ to a complex OP, a macroscopic wave function,
as it appears in a superfluid or superconductor by defining
$\psi_0=\phi_0^x - \mbox{i} \phi_0^y$. Apart from demonstrating that
all these systems belong to the same  the static universality class it also
is of practical advantage in the dynamic calculation.

The critical dynamics of
relaxing OPs coupled to a diffusing secondary density is governed by
the following equations of motion \cite{dohmjanssen77}
(there the complex OP $\psi_0$ was used):
\begin{eqnarray}
\label{dphiperp} \frac{\partial \phi_{\perp 0}^\alpha}{\partial
t}&=&-\mathring{\Gamma}^\prime_\perp \frac{\delta {\mathcal
H}}{\delta \phi_{\perp 0}^\alpha}
+\mathring{\Gamma}^{\prime\prime}_\perp \epsilon^{\alpha\beta z}
\frac{\delta {\mathcal H}}{\delta \phi_{\perp 0}^\beta} \nonumber \\
&+&\mathring{g}\ \epsilon^{\alpha\beta z} \phi_{\perp
0}^\beta\frac{\delta {\mathcal H}} {\delta
m_0}+\theta_{\phi_\perp}^\alpha \ ,
\end{eqnarray}
\begin{eqnarray}
\label{dphipar} \frac{\partial \phi_{\|0}}{\partial
t}&=&-\mathring{\Gamma}_\| \frac{\delta {\mathcal H}}{\delta
\phi_{\|0}}+\theta_{\phi_\|}
\, , \\
\label{dmpar} \frac{\partial m_0}{\partial
t}&=&\mathring{\lambda}\nabla^2 \frac{\delta {\mathcal H}}{\delta
m_0}+ \mathring{g}\ \epsilon^{z\alpha\beta}\phi_{\perp 0}^\alpha
\frac{\delta {\mathcal H}}{\delta\phi_{\perp 0}^\beta}
 +\theta_m \, ,
\end{eqnarray}
with the Levi-Civita symbol $\epsilon^{ijk}$. Here
$\alpha,\beta=x,y$ and the sum over repeated indices is implied.
Combining the kinetic coefficients of the OP to  a complex quantity,
$\mathring{\Gamma}_\perp=\mathring{\Gamma}_\perp^\prime
+\mbox{i}\mathring{\Gamma}_\perp^{\prime\prime}$, the imaginary part
constitutes a precession term created by the renormalization
procedure even if it is absent in the background. The kinetic
coefficient $\mathring{\lambda}$ and the mode coupling
$\mathring{g}$ are real. The stochastic forces
$\vec{\theta}_{\phi_\perp}$, $\vec{\theta}_{\phi_\|}$ and $\theta_m$
fulfill Einstein relations
\begin{eqnarray} \label{thetaperp}
\langle\theta_{\phi_\perp}^\alpha(x,t)\ \theta_{\phi_\perp}^\beta
(x^\prime,t^\prime)\rangle \!\!\!&=&\!\!\!
2\mathring{\Gamma}^\prime_\perp\delta(x-x^\prime)\delta(t-t^\prime)\delta^{\alpha\beta}
\ , \\
\label{thetapara} \langle\theta_{\phi_\|}(x,t)\
\theta_{\phi_\|}(x^\prime,t^\prime) \rangle \!\!\!&=&\!\!\!
2\mathring{\Gamma}_\|\delta(x-x^\prime)\delta(t-t^\prime)
\ , \\
\label{thetam} \langle\theta_m(x,t)\ \theta_m(x^\prime,t^\prime)
\rangle \!\!\!&=&\!\!\!
-2\mathring{\lambda}\nabla^2\delta(x-x^\prime)\delta(t-t^\prime) \ .
\end{eqnarray}

The reversible terms in these dynamic equations  for the OP components and the
conserved density have been derived by using generalized Poisson brackets for the
spin components  defining the  staggered magnetization and the $z-$ component of
the magnetization (for more details see \cite{review} and the literature cited there).
An exception constitutes the term with $\mathring{\Gamma}^{\prime\prime}_\perp$
which appears due to the renormalization procedure and the nonzero asymmetric
coupling $\mathring{\gamma_\perp}$. Similar dynamic equations may also appear
in the other systems where biconical 
phases are observed. The superfluid transition is described by model F and also
for the superconducting transition this universality class has been suggested
\cite{Lidmar98,Nogueira05} although no explicit derivation based on the methods
used here has been performed.

Applying the renormalization procedure using minimal subtraction
scheme \cite{bauja76} we find the flow equations for the time scale
ratios of the renormalized kinetic coefficients and the mode
coupling between  the perpendicular OP components and the
magnetization. We define  time scale ratios by the ratios of the
kinetic coefficients of the OPs and the secondary density
 $w_\perp\equiv\frac{\Gamma_\perp}{\lambda}$, $w_\|\equiv\frac{\Gamma_\|}{\lambda}$,
 as well as the ratios between the relaxation rates of the two
OPs $v\equiv\frac{\Gamma_\|}{\Gamma_\perp}=\frac{w_\|}{w_\perp}$,
$v_\perp\equiv\frac{\Gamma_\perp}{\Gamma_\perp^+}=\frac{w_\perp}{w_\perp^+}$,
and the mode coupling parameters $f_\perp\equiv
g/\sqrt{\Gamma_\perp^{\prime}\lambda}$ or $F=g/\lambda$. For these
dynamic parameters we obtain  the flow equations
\begin{eqnarray}
\label{dwdl} l\frac{d
w_\perp}{dl}&=&w_\perp\left(\zeta_{\Gamma_\perp}-\zeta_\lambda\right)
\ , \quad
l\frac{d w_\|}{dl}=w_\|\left(\zeta_{\Gamma_\|}-\zeta_\lambda\right) \ , \\
\label{df1dl} l\frac{d
f_\perp}{dl}&=&-\frac{f_\perp}{2}\left(\epsilon+\zeta_\lambda-2\zeta_m
+\Re\left[\frac{w_\perp}{w_\perp^\prime}\
\zeta_{\Gamma_\perp}\right]\right) \, ,
\end{eqnarray}
where $l$ is the RG flow parameter and the $\zeta_\lambda$-function
is obtained by the renormalization procedure as
\begin{equation}\label{zetalad}
\zeta_\lambda=\frac{1}{2}\gamma_\perp^2+\frac{1}{4}\gamma_\|^2-
\frac{f_\perp^2}{2}\Big(1+{\cal Q}\Big) \ .
\end{equation}
The function ${\cal Q}\equiv {\cal Q}(\gamma_\perp,w_\perp,F)$
contains all higher order contributions beginning with two loop
order and is identical to the corresponding function in model F (see
(A.28) and (A.29) in \cite{review}). We obtain the $\zeta$-function
for the perpendicular kinetic coefficient  $\Gamma_\perp$ as
\begin{eqnarray}\label{zetagammape}
\zeta_{\Gamma_{\perp}}=\zeta_{\Gamma_{\perp}}^{(A)}\big(\{u\},v_\perp,v\big)
+
\frac{D_\perp^2}{w_\perp(1+w_\perp)} \nonumber \\
-\frac{2}{3}\frac{u_\perp D_\perp}{w_\perp(1+w_\perp)}\ A_\perp
-\frac{1}{2}\frac{D_\perp^2}{w_\perp^2(1+w_\perp)^2}\ B_\perp  \nonumber \\
-\frac{1}{2}\frac{\gamma_\|D_\perp}{1+w_\perp}\left(\frac{u_\times}{3}
+\frac{1}{2}\frac{\gamma_\|D_\perp}{1+w_\perp}\right)X_\perp   \, ,
\end{eqnarray}
where we have introduced the coupling $D_\perp\equiv
w_\perp\gamma_\perp-\mbox{i}F$. The functions $A_\perp\equiv
A_\perp(\gamma_\perp,\Gamma_\perp,w_\perp,F)$, $B_\perp\equiv
B_\perp(\gamma_\perp,\Gamma_\perp,w_\perp,F)$ are identical to
eqs.~(A.25), (A.26) in \cite{review}. $X_\perp$ is defined as
\begin{equation}\label{xperp}
X_\perp\equiv
1+\ln\frac{2v}{1+v}-\left(1+\frac{2}{v}\right)\ln\frac{2(1+v)}{2+v}
\, ,
\end{equation}
$\zeta_{\Gamma_{\perp}}^{(A)}\big(\{u\},v_{\perp},v\big)$ is the
$\zeta$-function of the perpendicular relaxation $\Gamma_\perp$ in
the biconical model A, but now with a complex kinetic coefficient
$\Gamma_\perp$
\begin{eqnarray}\label{zetagaape}
\zeta_{\Gamma_{\perp}}^{(A)}\big(\{u\},v_{\perp},v\big)=
\frac{u_\perp^2}{9}\Big(2\ln\frac{2}{1+\frac{1}{v_\perp}} \nonumber \\
+ (2+v_\perp)\ln\frac{\left(1+\frac{1}{v_\perp}\right)^2}
{1+2\frac{1}{v_\perp}}-\frac{1}{2}\Big) \nonumber \\
+\frac{u_\times^2}{36}\left(\ln\frac{(1+v)^2}{v(2+v)}+\frac{2}{v}\ln\frac{2(1+v)}
{2+v}-\frac{1}{2}\right) \, .
\end{eqnarray}
The dynamic $\zeta$-function of the parallel relaxation kinetic
coefficient $\Gamma_\|$ is obtained as
\begin{eqnarray}\label{zetagammapa}
\zeta_{\Gamma_{\|}}=\bar{\zeta}_{\Gamma_{\|}}^{(C)}(u_\|,\gamma_\|,w_\|)
-\frac{1}{2}\frac{w_\|\gamma_\|}{1+w_\|}\Bigg[\left(\frac{2}{3}u_\times
{+}\frac{w_\|\gamma_\|}{1{+}w_\|}\gamma_\perp\right) \nonumber \\
\times \Re\Big[\frac{T_1}{w_\perp^\prime}\Big]
-\frac{\gamma_\|F}{2(1{+}w_\|)} \Im\Big[\frac{T_2}{w_\perp^{\prime
2}}\Big]\Bigg]
+\zeta_{\Gamma_{\|}}^{(A)}\big(\{u\},v_\perp,v\big) \, . \nonumber \\
\end{eqnarray}
$\bar{\zeta}_{\Gamma_{\|}}^{(C)}(u_\|,\gamma_\|,w_\|)=
\zeta_\Gamma(u_\|,\gamma_\|,\Gamma_\|,w_\|)-\zeta_\Gamma^{(A^\star)}(u_\|,\Gamma_\|)$,
where the functions on the right hand side are defined by (A.8) and
(A.9) for $n=1$ in \cite{review}. The functions $T_1$ and $T_2$ are
defined as
\begin{eqnarray}\label{T1}
T_1\equiv&&\!\!\!\!D_\perp\Bigg[1+\ln\frac{1+\frac{1}{v_\perp}}{1+v}  \nonumber\\
&&\!\!\!\!-\left(v{+}\frac{1}{v_\perp}(1{+}v)\right)
\ln\frac{(1{+}v)\left(1{+}\frac{1}{v_\perp}\right)}{v{+}\frac{1}{v_\perp}(1{+}v)}\Bigg]
\ ,
\end{eqnarray}
\begin{eqnarray}\label{T2}
T_2\equiv w_\perp^+D_\perp\Bigg[(1+v_\perp)v
-\ln\frac{1+\frac{1}{v_\perp}}{1+v} \nonumber \\
-\left(v{+}\frac{1}{v_\perp}(1{+}v)\right)\big(v{+}v_\perp(1{+}v)\big)
\ln\frac{(1{+}v)\left(1{+}\frac{1}{v_\perp}\right)}{v{+}\frac{1}{v_\perp}(1{+}v)}\Bigg]
\ ,
\end{eqnarray}
and $\zeta_{\Gamma_{\|}}^{(A)}\big(\{u\},v_{\perp},v\big)$ is the
$\zeta$-function of the kinetic coefficient of the parallel
relaxation in the biconical model A. With a complex $\Gamma_\perp$
it reads
\begin{eqnarray}\label{zetagaapa}
\zeta_{\Gamma_{\|}}^{(A)}\big(\{u\},v_{\perp},v\big)=
\frac{u_\|^2}{4}\left(\ln\frac{4}{3}-\frac{1}{6}\right)  \nonumber \\
+\frac{u_\times^2}{18}\Bigg(\ln\frac{(1{+}v)\left(\frac{1}{v_\perp}{+}v\right)}
{v{+}\frac{1}{v_\perp}(1{+}v)}
+vv_\perp\ln\frac{\left(1{+}\frac{1}{v_\perp}\right)\left(\frac{1}{v_\perp}{+}v\right)}
{v{+}\frac{1}{v_\perp}(1{+}v)}   \nonumber \\
+v\ln\frac{\left(1{+}\frac{1}{v_\perp}\right)(1{+}v)}{v{+}\frac{1}{v_\perp}(1{+}v)}-
\frac{1}{2}\Bigg)\, .
\end{eqnarray}

\begin{table}
\caption{Two loop FP values of the mode coupling $f_\perp$,  the
ratios $q=w_\|/w_\perp^\prime$,
$s=w_\perp^{\prime\prime}/w_\perp^\prime$ and the dynamic exponents
in the subspace $w_\|=0$, $w_\perp=0$ with finite value of
$v=q/(1+is)$ for the static biconical ${\cal B}$ and Heisenberg
${\cal H}$ FPs. For comparison we add the FP values for the
exponents that govern critical dynamics at magnetic fields below and
above the multicritical point. These are described by  model C at
$n=1$ and model F at $n=2$.} \label{tableI}
\begin{center}
\begin{tabular}{llllll} \hline \hline
       & $f_\perp^\star$ & $q^{\star}$ &$s^{\star}$ &$z_{\rm OP}$ &  $z_m$ \\ \hline
${\cal B}$ & $1.232$ & $1.167\cdot 10^{-86}$ & $0$ & $2.048$   & $1.131$ \\
${\cal H}$ &  $1.211$ &  $3.324\cdot 10^{-8}$ & $0$ & $2.003$ & $1.542$ \\
${\cal B}$ & $1.232$ & $2.51\cdot 10^{-782}$ & $0.705$ & $2.048$     & $1.131$ \\
${\cal H}$ &  $1.211$ &  $3.16\cdot 10^{-66}$ & $0.698$ & $2.003$   & $1.542$ \\
C   \cite{C} & - &  -  & -& $2.18$ & $2.18$\\
F \cite{F}&  $0.83$ & - & - & $\sim 1.5$  & $\sim 1.5$\\
\hline \hline
\end{tabular}
\end{center}
\end{table}

In order to find the FP values of the time-scale ratios and the mode
coupling the right hand sides of eqs.~(\ref{dwdl})-(\ref{df1dl})
have to be zero. If the FP value of the mode coupling $f_\perp$
would be zero one obtains the FP values of the time ratios of model
C discussed in \cite{partIII}. However this FP is unstable. If the
FP value of $f_\perp$ is nonzero then due to the logarithmic terms
in the $\zeta$-functions both OP have to have the same time scales
i.e. a finite nonzero FP value $v^\star$. This is only possible
either for nonzero finite FP values of $w_\perp$ and $w_\|$ or when
both of these FP values are zero. No finite FP values for $w_\perp$ and $w_\|$
have been found. In the other case the {\it
approach} to zero of both time scales has to be the same. Therefore
the approach to the multicritical dynamic FP is described by the
flow in the limit $w_\perp\to 0$, $w_\|\to 0$ and $v$ finite
(asymptotic subspace). The flow  in the complete dynamic parameter
space and in this asymptotic subspace will be discussed afterwards.

The $\zeta$-function for the perpendicular OP relaxation might be
complex,
$\zeta_{\Gamma_\perp}=\zeta^\prime_{\Gamma_\perp}+i\zeta^{\prime\prime}_{\Gamma_\perp}$.
In order to obtain the usual asymptotic power laws for the
relaxation coefficients $\Gamma_\|$ and $\Gamma_\perp$ the FP value
of the imaginary part $\zeta^{\prime\prime\star}$ has to be zero. In
consequence the asymptotic flow of the real and imaginary parts of
$v$  is governed by the same exponent
$\zeta^{\prime\star}_{\Gamma_\perp}-\zeta^\star_{\Gamma_\|}$.

\begin{figure}
\onefigure{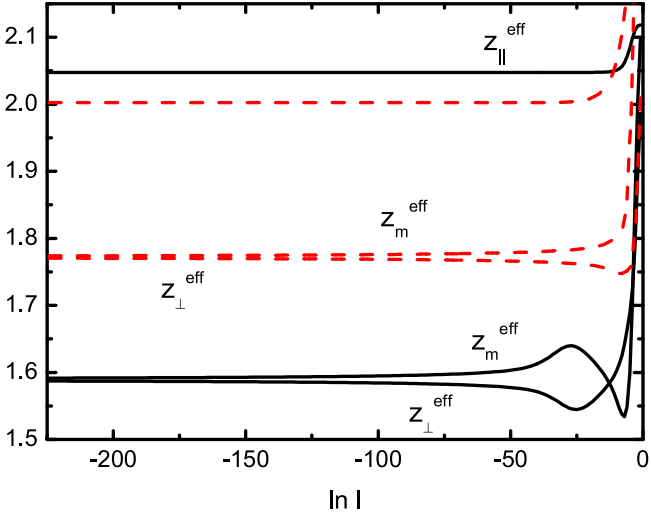} \caption{Effective dynamic exponents in the
background using the flow equations  (\ref{dwdl}),(\ref{df1dl}) in
the complete dynamical parameter space. The static values are taken
for the Heisenberg FP (dashed curves) and for the biconical FP
(solid curves).} \label{zeffback}
\end{figure}

If the FP value of the mode coupling $f$ is different from zero and
finite one has from eq.~(\ref{df1dl})
$\varepsilon+\zeta_{\Gamma_\perp}^{\prime\star}+\zeta^{(d)\star}_\lambda=0$
and the relation \cite{dohmmulti83} between dynamical and static
critical exponents $z_{\perp}+z_m=2\frac{\phi}{\nu}$ (here the $z$
exponents govern  the corresponding scaling times and  $\phi$ and
$\nu$ are the crossover and
 correlation length exponents). The dynamical exponents are defined as
$z_o=2+\zeta_o^\star$ with $o=\perp, \, \|, \, m$. Because $v^\star$
is  finite and nonzero $z_{\perp}=z_{\|}\equiv z_{\rm OP}$. This
means that strong scaling  with respect to the OPs, the  components
of the staggered magnetizations, but weak scaling with respect to
the conserved density, the magnetization $m$, holds since $z_m\neq
z_{\rm OP}$.

\begin{figure}[t]
             \onefigure{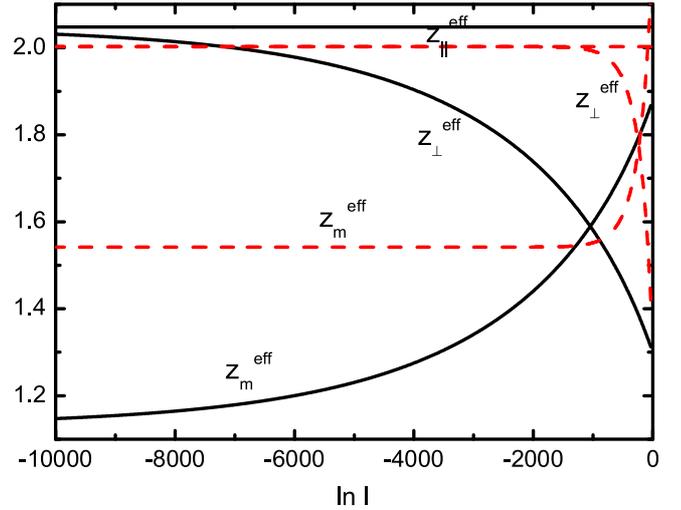}
\caption{Effective dynamic exponents in the asymptotic subspace
$w_\|=w_\perp=0$ and $v \equiv w_\|/w_\perp\neq 0$ and finite.
Dashed and solid curves as in Fig. \ref{zeffback}.}
 \label{zetasub}
\end{figure}

The two loop order values of the dynamic exponents together with the 
FP values of the time scales and the mode coupling are presented in 
table~\ref{tableI}. For the model of the three-dimensional 
anisotropic antiferromagnet under consideration, the biconical FP 
${\cal B}$ ($u^\star_\perp \neq u^\star_\| \neq u^\star_\times$) has 
been shown to be stable. It governs the static critical behaviour in the 
complete space of couplings (see e.g. \cite{partI}). Substituting 
their two-loop values obtained in \cite{partI} within generalized 
Pad\'e-Borel resummation technique \cite{resummation} into the flow 
equations (\ref{dwdl})--(\ref{df1dl}), we get two dynamical FPs, 
their coordinates are given in the first and third row of 
table~\ref{tableI}. Although  two different dynamical FPs are found
(with zero and nonzero $s^\star$) this difference does not lead to a
change in the corresponding FP values of the dynamical exponents.
This is because both FPs have extremely small but different
$q^\star$.

For comparison we have included, besides the biconical FP ${\cal B}$ 
describing tetracritical behavior, the isotropic Heisenberg FP ${\cal 
H}$  ($u^\star_\perp = u^\star_\| = u^\star_\times$) describing 
bicritical behavior. This FP is only reached in the subspace of the 
static couplings that lie in its attraction region (see fig.~3 in 
\cite{partI}). Again, as in the case of the biconical FP  ${\cal B}$, we obtain 
two dynamical FPs with coordinates given in the second and fourth 
line of table~\ref{tableI}. The numerical values of the dynamical 
exponents are practically equal in the different dynamical FPs ${\cal 
H}$.  We further quote
in table~\ref{tableI} the dynamical critical exponents on the two
phase transition lines below and above the multicritical point,
which are given by model C  and model F respectively.

The FP value of $v$  is extremely small and therefore  in the
physical accessible region one cannot prove strong scaling for the
OP components. Indeed in the non-asymptotic region the dynamic
parameters are described by the flow equations
(\ref{dwdl}),(\ref{df1dl}) and from these dependencies the effective
dynamic exponents can be calculated. The result is shown in
fig.~\ref{zeffback}. The static parameters have been set already to
their FP values and therefore the starting values of the effective
exponents are different from  $z=2$. It turns out that the prefactor
of the $\ln v$-terms in eqs.~(\ref{zetagammape}) and (\ref{xperp}),
 which
drive the flow of the dynamic parameters into the asymptotic
subspace is reduced and the flow is almost like in one loop order.
Therefore
 weak scaling with $z^{eff}_{\|}\sim 2.04$ and
$z^{eff}_{\perp}\sim z^{eff}_m\sim 1.6$ is observed.\footnote{Even
for flow parameters $\ln l\sim -10^6$ no visible changes in the
values of the different $z^{eff}$ occur although the dynamic
parameters change. Thus the asymptotic subspace is not reached for
these extremely small values of $l$.}

The approach of the effective dynamical exponents in the asymptotic
subspace $w_\perp=w_\|=0$ and $v$ finite  to their biconical FP
values is shown in fig.~\ref{zetasub}. The background behavior is
dominated by a behavior corresponding for the perpendicular
components by model F  and for the parallel components by model A
with a finite value Re($v$) whereas  Im($v$) is almost zero.
Therefore for the biconical case even for flow parameter
values $\ln l\sim -10^4$  the two effective exponents do not reach  
their asymptotics:
$z^{eff}_{\perp}<z_{\rm OP}$ and $z^{eff}_m>z_m$. This is different
for the Heisenberg case where the FP values of the dynamical
exponents are reached (see dashed curves in fig.~\ref{zetasub}).

Although from our calculation we conclude that the
asymptotics (strong scaling) would be unobservable effective exponents as
described in fig.~\ref{zeffback} are observable. 
The complete two loop calculation allowed us to calculate not only the 
values of the dynamic FP but also the effective exponents which are the quantities 
governing the behavior of the transport coefficients, i.e. the relaxation and diffusion 
coefficient of the staggered magnetization and magnetization respectively.
It is well known that near a dynamical stability boundary separating a strong scaling FP
with a finite timescale ratio from a weak scaling FP with a vanishing time scale ratio 
also small dynamic transient exponents appear and effective critical behavior is observed.
The case where the OPs have the component values $n=2$ and $n=1$ is located near the
stability boundary between  the biconical and decoupling FP as has been demonstrated
in \cite{partI}. For the decoupling FP the time scales of the two OPs
scale differently and weak scaling is expected. The effective values of the dynamical
critical exponents (starting from different initial conditions) are driven to almost
stationary values. These might be measured in neutron scattering experiments.

A natural question concerns the reliability of numerical predictions 
for the observables obtained in our study. In particular, how will 
an increase of the order of the perturbation theory influence the 
numerical estimates. An estimate can be given by comparing results 
obtained in different perturbation theory (loop) orders. This can be 
done for the static part of the RG functions, which are currently 
known with a record five loop accuracy \cite{high_order}. As it was 
demonstrated in Ref. \cite{partI}, the two loop approximation we 
consider here refined by the resummation is enough to catch the main 
features of the static phase transition (FP stability and respective 
universality classes) as well as to give reliable estimates for the 
observable quantities that govern the phase transition (leading 
exponents and corrections to scaling). It is well known, that the 
dynamic RG calculations are technically much more complicated as 
static ones. In particular, no higher orders of the perturbation 
theory are known for the model we consider here. However, as known from 
the previous experience in the RG description of dynamical criticality 
\cite{review} we expect also in this case that the two loop calculation
captures the essential dynamical properties, namely to be effectively in 
a weak scaling situation.

The comparison between experiment and theory for multicritical behavior is much
less developed in
dynamics than in statics (see e.g. the situation at the tricritical point
\cite{tricrit}).
But exploring even the critical dynamics along the two transition lines is of
interest since
the OP and the condserved density are experimentally accessible. Also  
computer simulation might be considered for a comparison \cite{landau08} and explicit
theoretical results are worthwhile for the interpretation of the numerical results.
Not only exponents are necessary for a careful 
interpretation but also the calculation of the dynamic structure factors are of interest.
This also concerns the neutron scattering experiments. The effective values of the timescale
ratios, known from flow equations (like (\ref{dwdl})), enter the  shape functions and may
change their shape from a Lorentzian considerably.

\acknowledgments This  work was supported by the FWF under Project
No. P19583-N20.


\begin{thebibliography}{99}

\bibitem{liufisher72}
 \Name{Liu K.-S. \and Fisher M. E.}
  \REVIEW{J. Low Temp. Phys.} {10} {1972} {655}.

\bibitem{chang04}
 \Name{Kim E. \and  Chan M. H. W.}
 \REVIEW{Nature} {427} {2004} {225}.

\bibitem{matsuda70}
 \Name{Matsuda H. \and Tsuneto T.}
 \REVIEW{Prog. Theor. Phys. Suppl.} {46} {1970} {411}.

\bibitem{knf76}
 \Name{Kosterlitz J. M.,  Nelson D. \and  Fisher M. E.}
 \REVIEW{Phys. Rev. B} {13} {1976} {412}.

\bibitem{partI}
 \Name{Folk R.,  Holovatch Yu. \and Moser  G.}
 \REVIEW{Phys. Rev. E} {78} {2008} {041124}.
 
\bibitem{shapira}
  \Name{Shapira Y.}
  \Book{Multicritical Phenomena (NATO ASI Series B)}
  \Editor{R. Pynn \and A. Skjeltrop}
  \Vol{106}
  \Publ{Plenum, New York}
  \Year{1983}
  \Page{35}. 

\bibitem{mukamel76}
\Name{Mukamel D.}
\REVIEW{Phys. Rev. B} {14} {1976} {1303}

\bibitem{aharony06}
\Name{Aharony A.} 
\REVIEW{J. Stat. Phys.} {110} {2003} {659}

\bibitem{review}
 \Name{Folk R. \and Moser G.}
 \REVIEW{J. Phys. A: Math. Gen.} {39} {2006} {R207}.


\bibitem{seabra10}
 \Name{Seabra L. \and  Shannon N.}
Phys. Rev. Lett., to be published (arXiv:1003.3430v2).

\bibitem{partII}
 \Name{Folk R.,  Holovatch Yu. \and  Moser G.}
 \REVIEW{Phys. Rev.E} {78} {2008} {041125}.

\bibitem{dohmjanssen77}
 \Name{Dohm V. \and  Janssen H.-K.}
 \REVIEW{Phys. Rev. Lett.} {39} {1977} {946}; \REVIEW{J. Appl. Phys.} {49} {1978}
 {1347}.

\bibitem{dohmKFA}
 \Name{Dohm V.} Kernforschungsanlage J\"ulich Report No. 1578, 1979
(unpublished).

\bibitem{dohmmulti83}
  \Name{Dohm V.}
  \Book{Multicritical Phenomena (NATO ASI Series B)}
  \Editor{R. Pynn \and A. Skjeltrop}
  \Vol{106}
  \Publ{Plenum, New York}
  \Year{1983}
  \Page{81}.

\bibitem{huber7476}
 \Name{Huber D. L.}
 \REVIEW{Phys. Lett. A} {49} {1974} {345};
 \Name{Huber D. L. \and  Raghavan R.}
 \REVIEW{Phys. Rev. B} {14} {1976} {4068}.
 
\bibitem{Lidmar98}
 \Name{Lidmar J. L, Wallin M., Wengel C., Girvin S. M., \and
Young A. P.} 
\REVIEW{Phys. Rev. B} {58} {1998}   {2827}.

\bibitem{Nogueira05} 
\Name{Nogueira F. S. \and  Manke D.} 
\REVIEW{Phys. Rev. B} {72} {2005} {014541}.

\bibitem{bauja76}
 \Name{Bausch R.,  Janssen H.K. \and  Wagner H.}
 \REVIEW{Z. Phys. B} {24} {1976} {113}.

\bibitem{partIII}
 \Name{Folk R.,  Holovatch Yu. \and  Moser G.}
 \REVIEW{Phys. Rev. E} {79} {2009} {031109}.

\bibitem{C}
 \Name{Folk R. \and Moser G.}
 \REVIEW{Phys. Rev. E} {69} {2004} {036101}.

\bibitem{F}
 \Name{Dohm V.}
 \REVIEW{Phys. Rev. B} {44} {1991} {2697}.
 
 \bibitem{resummation} 
 \Name{Holovatch Yu.,  Blavats'ka V.,  Dudka M.,  von Ferber C. \and Folk R.  \and Yavors'kii T.}
 \REVIEW{Int. J. Mod. Phys. B} {16}{2002}{4027}. 

\bibitem{high_order} 
 \Name{Calabrese P., Pelissetto A., \and  Vicari E.}
 \REVIEW{Phys. Rev. B} {67}{2003}{054505}.
 
 \bibitem{tricrit}
 \Name{Folk R. \and Moser G.}
 \REVIEW{J. Low Temp. Phys.} {150}{2007}{689}.
 
 \bibitem{landau08}
 \Name{Shan-Ho Tsai \and   Landau D.P.}
 \REVIEW{Computing in Science \& Engineering}
 {10}{2008}{72}
 


\end{thebibliography}
\end{document}